\documentstyle[preprint,eqsecnum,aps]{revtex}
\begin{document}
\draft
\preprint{UW-Madison}
\title{Observation of a van Hove Singularity\\
in $Bi_{2}Sr_{2}Ca_{1}Cu_{2}O_{8+x}$ with Angle Resolved Photoemission}
\author{Jian Ma,$^{a}$ C.Quitmann,$^{a}$ R.J.Kelley,$^{a}$\\
P.Alm\'{e}ras,$^{b}$ H.Berger,$^{b}$ G. Margaritondo,$^{b}$
and M.Onellion$^{a}$}
\address{$^{a}$Department of Physics and Applied Superconductivity Center,
University of Wisconsin-Madison,
Madison, WI 53706\\
$^{b}$Institut de Physique Appliqu\'{e}e, Ecole Polytechnique F\'{e}d\'{e}rale,
CH-1015 Lausanne, Switzerland}
\date{\today}
\maketitle

\begin{abstract}
We have performed high energy resolution angle-resolved photoemission
studies of the normal state band structure of oxygen overdoped
$Bi_{2}Sr_{2}Ca_{1}Cu_{2}O_{8+x}$. We
find that there is an extended saddle point singularity in the density of
states along $\Gamma-\bar{M}-Z$ direction. The data also indicate that
there is an asymmetry in the Fermi surface for both the
$\Gamma-\bar{M}-Z$ and perpendicular directions.
\end{abstract}
\pacs{PACS numbers: 73.20.At, 79.60.-i, 74.25.Jb, 74.72.Hs}

\narrowtext

\section{Introduction:}
\label{sec:level1}
The possibility of a van Hove singularity has been of considerable interest
since it was proposed as a means of enhancing the superconducting transition
temperature, $T_{c}$.\cite{Newns,Hirsch,Labbe,Friedel}
The idea is that the tendency toward
superconductivity in a two-dimensional system can be enhanced when the
Fermi level lies at or close to the energy of a logarithmic van Hove
singularity (VHS) in the density of states.\cite{Hirsch,Friedel} Several
investigators have studied such a
mechanism for raising the superconducting transition temperature
$T_{c}$\cite{Markiewicz,Abrikosov,Newns2,Takacs,Radtke} and there
has been indirect
experimental evidence indicating the possibility of a van Hove
singularity.\cite{Tsuei,Pattnaik} More
recently, the first direct spectroscopic observations of a van Hove
singularity, in the $YBa_{2}Cu_{3}O_{7-x}$ and $YBa_{2}Cu_{4}O_{7-x}$ systems
have been reported.\cite{Gofron,Liu} Gofron et.al. show that
there is an extended saddle point singularity along the $\Gamma-Y$ direction in
reciprocal space, centered at the Y-point.\cite{Gofron}
 This report motivated us to study
the $Bi_{2}Sr_{2}Ca_{1}Cu_{2}O_{8+x}$ system.

Our main result is that there is indeed an extended saddle point type of van
Hove singularity in the oxygen overdoped $Bi_{2}Sr_{2}Ca_{1}Cu_{2}O_{8+x}$
system. We have also
obtained two additional results: the electronic band structure is asymmetric
along both the $\Gamma-\bar{M}-Z$ and perpendicular directions, and there
is no evidence for an electronic state just above the Fermi energy along the
$\Gamma-\bar{M}-Z$ direction.

\section{Experimental:}
The experiments were performed using the four meter normal incidence
monochromator at the Wisconsin Synchrotron Radiation Center in Stoughton, WI.
The beamline provides highly ($>$ 95\%) linearly polarized light with
the electric vector in the horizontal plane and with photon energy
resolution better than 10 meV. The angle-resolved photoemission chamber
includes
a reverse-view low energy electron diffraction (LEED) optics used to orient
the sample {\it in situ} after cleaving. The electron
energy analyzer
is a 50 mm VSW hemispherical analyzer mounted on a two axis goniometer, with
an acceptance full angle of $2^\circ$. The base pressure is
$6 \times 10^{-11}$ torr. The incidence angle between the photon Poynting
vector and surface normal was $45^\circ$ unless otherwise noted.

The single crystal samples were annealed in 1 atmosphere of oxygen at
530 $^\circ$C for 20 hours. Figure\ \ref{AC}
illustrates the results of
AC susceptibility measurements. Note particularly
that $T_{c}$ is 83K and the 10-90\% temperature width is 1.3K. The crystals
were
characterized using four-point resistivity, X-ray diffraction, and transmission
electron microscopy. The samples were transferred from a load lock
chamber with a
base pressure of $5 \times 10^{-9}$ torr to the main chamber, and
cleaved at 30K in a vacuum of $6-8 \times 10^{-11}$ torr. The sample holder
includes the capability to rotate the sample {\it in situ} about the surface
normal, at low temperatures, for precision alignment with respect to the
photon electric field. To measure the normal state electronic band structure,
the temperature was raised to 95K, above $T_{c}$ . The stability of
the temperature was $\pm1K$. We obtained a Fermi edge reference by using the
spectra of freshly deposited gold films. We have obtained an energy
resolution of 15 meV using 1 eV pass energy. For the present study, the overall
energy resolution employed was 35 meV unless otherwise stated.

For a quasi-two-dimensional system such as $Bi_{2}Sr_{2}Ca_{1}Cu_{2}O_{8+x}$
the initial state of the
electron can be determined by measuring the component of the electron momentum
parallel to the sample surface ({\bf k}$_{//}$). By measuring the energy
distribution curves (EDC's) for different directions
($\theta$, $\phi$) of the emitted photoelectron relative to the surface
normal, the {\bf k}$_{//}$ of the
initial state is derived from the relation:
{\bf k}$_{//}=0.512$\AA$^{-1} \sqrt{E_{kin}}(\sin\theta \cos\phi \hat{k_{x}} +
\sin\phi \hat{k_{y}})$, where $E_{kin}$ is the kinetic energy of measured
photoelectrons in the unit of eV, $\hat{k_{x}}$ and $\hat{k_{y}}$ denote
unit vectors along two Cu-O bond axis ($\Gamma-\bar{M}$ directions).
The binding energy of the state is determined by using the Fermi edge of
the gold film as a reference.

\section{Results and Discussion:}
We first oriented the sample so that the $\Gamma-\bar{M}-Z$ direction was in
the
horizontal plane. We measured the dispersion of the normal state quasiparticle
band along the $\Gamma-\bar{M}-Z$ horizontal axis. Figure\ \ref{GammaMZ}(a)
and (b) illustrate the spectra obtained using an energy resolution of
(a) 55 meV and (b) 35 meV, and a photon energy of (a) 21 eV and (b) 25 eV.
The emission angle of the photoelectrons for each
spectrum is noted in the figure.

A band developes and is visible for
$8^\circ$  off normal
at a binding energy of $335\pm5$ meV. The band disperses toward the Fermi
energy and, at $\theta = 20^\circ$, is at a binding energy of 0-15 meV. The
$\bar{M}$ point at the photon energy of 25 eV is at $21^\circ$. For this
polarization (even symmetry) there is no indication of a Fermi level
crossing,\cite{Olson}
 consistent with reports of Dessau et.al.\cite{Dessau}

Between
$\theta = 20^\circ-32^\circ$ the quasiparticle band remains at a binding
energy of 0-15 meV. Starting at $34^\circ$ the quasiparticle band disperses
to higher binding energy, reaching a binding energy of 120 meV at an angle of
$40^\circ$. The dispersion curve is illustrated in Fig.\ \ref{GammaMZ}(c).
The data in Fig.\ \ref{GammaMZ} illustrate what is, to the best of our
knowledge, the first observation of
the quasiparticle dispersion near the Z-point which, for this photon energy,
is $43^\circ$. All of the above results on
the quasiparticle dispersion along the
$\Gamma-\bar{M}-Z$ direction were confirmed using a photon energy of 21 eV.
This behavior, with the quasiparticle band near the Fermi energy for an
extended portion of the $\Gamma-\bar{M}-Z$ line, is consistent with an
extended saddle point singularity.\cite{Gofron} Note also that the dispersion
is not symmetric with respect to the $\bar{M}$ point; we return later to this
point.

The other crucial issue of the band structure topology is the behavior of the
quasiparticle band along cuts perpendicular to the $\Gamma-\bar{M}-Z$
direction. To obtain definitive information, we made four cuts, at
angles $14^\circ$, $16^\circ$, $18^\circ$ and $20^\circ$ along the
$\Gamma-\bar{M}-Z$ direction. In Figs.\ \ref{X-M-Y}-\ref{cut14}, we
illustrate the results obtained with 25 eV photon energy. We also confirmed the
results of Figs.\ \ref{X-M-Y}-\ref{cut14} by using a photon energy of 21 eV
(data not shown). Fig.\ \ref{X-M-Y}(a) and (b) illustrate the results
for the cut at $\theta = 20^\circ$, very close
to the $\bar{M}$ point along the $\Gamma-\bar{M}-Z$ line.
Fig.\ \ref{X-M-Y}(a) illustrates the spectra along the $Y-\bar{M}-X$ line.
Along the $\bar{M}-X$ line, the intensity of the quasiparticle band
decreases monotonically with increasing angle. At $\phi = -5^\circ$ only the
background remains. The data indicate that the band disperses up through
the Fermi energy almost immediately as one moves from $\bar{M}$ along
the $\bar{M}-X$ line. By contrast, along the $\bar{M}-Y$ line, the band
first disperses slightly away from the Fermi surface, then exhibits a clear
Fermi surface crossing at $\phi = +7^\circ$. Figure\ \ref{X-M-Y}(b) illustrates
the dispersion along the $X-\bar{M}-Y$ direction, with Fermi surface
crossings as one moves away from the $\bar{M}$ point in either direction, but
the Fermi surface crossings in the $\bar{M}-X$ and $\bar{M}-Y$ directions
are not symmetric with respect to the $\Gamma-\bar{M}-Z$ line.

Another noteworthy point about Fig.\ \ref{X-M-Y}(a) is that we obtained data
all the way from the $\bar{M}$ point to the Y point. Notice that there is
only one Fermi surface crossing visible in the data. Since there is a part of
the Fermi surface around the Y-point, our data indicate that there is no
Bi-pocket around the $\bar{M}$ point along the
$\bar{M}-Y$ line.\cite{Massidda}

Figure\ \ref{cut18} illustrates the cut at $18^\circ$  parallel to the
$X-\bar{M}-Y$ line. The spectra in Fig.\ \ref{cut18}(a) exhibit an abrupt
reduction in the quasiparticle spectral feature intensity between
$\theta/\phi = 18^\circ/-4^\circ$ and $\theta/\phi = 18^\circ/-5^\circ$ in
going toward X, but the corresponding Fermi surface crossing in going
toward Y occurs between $\theta/\phi = 18^\circ/5^\circ$ and
$\theta/\phi = 18^\circ/6^\circ$. Fig.\ \ref{cut18}(b) illustrates the details
of the dispersion observed in Fig.\ \ref{cut18}(a). Note in particular that
the dispersion is again asymmetric about the $\Gamma-\bar{M}-Z$ line,
with a maximum binding energy of 60 meV at $\phi = +3^\circ$. Also, it is
noteworthy that for this cut, again, the quasiparticle state disperses
above the Fermi energy as one moves away from the $\Gamma-\bar{M}-Z$ line
in either direction.

We also studied a cut at $\theta = 16^\circ$ parallel to the $X-\bar{M}-Y$
line.
Figure\ \ref{cut16}(a) illustrates the spectra, and Fig.\ \ref{cut16}(b) the
resulting dispersion relation. For this cut, the dispersion is almost
symmetric, with a saddle point centered at $\phi = -1^\circ$. Note that
the quasiparticle state again disperses above the Fermi energy
as one moves away from the $\Gamma-\bar{M}-Z$ line in either direction. The
results of the cuts at $16^\circ$, $18^\circ$ and $20^\circ$ all indicate
that the Fermi surface topology is markedly different
around the X and the Y points.

Fig.\ \ref{cut14} illustrates a cut at $\theta = 14^\circ$ parallel to the
$X-\bar{M}-Y$ line. Note that the binding energy, 130 meV, along the
$\Gamma-\bar{M}-Z$ line for this cut places the quasiparticle state well
below the Fermi energy. For this reason, we would expect the behavior of the
quasiparticle state in this part of the Brillouin zone not to be strongly
involved in the superconducting properties. Note, however,
that we observe asymmetric  Fermi surface crossings along this cut at
$\theta/\phi = 14^\circ/-5^\circ$ and at $\theta/\phi = 14^\circ/+7^\circ$,
another indication that the shape of the Fermi surface around the X and the
Y points is different.

Figure\ \ref{FS} summarizes (a) our partial Fermi surface mapping and (b)
the shape of the Fermi surface near the $\Gamma-\bar{M}-Z$ direction.
 The mapping is
a purely experimental derivation; we have made no {\it a priori} assumptions
regarding the
symmetry of the Fermi surface. As noted above, we observe an asymmetric Fermi
surface shape with respect to the $\bar{M}$ point along the
$\Gamma-\bar{M}-Z$
line. The cause of this asymmetry is unknown; however, our observations are
consistent with our finding that  the $\Gamma-Y$ and $\Gamma-X$ lines are
inequivalent.\cite{Kelley} Note also that the shape of the Fermi surface is
as expected for an extended saddle point type of van Hove singularity.

Our results are different from those
reported by Ref.\ \onlinecite{Dessau} who, however, studied more lightly doped
samples (ours are overdoped with oxygen), and who assumed
tetragonal symmetry. Our data establish that for the overdoped samples,
the Fermi surface exhibits orthorhombic, rather than tetragonal, symmetry.
Such a result is consistent with X-ray and transmission electron microscopy
studies of our crystals, which both indicate orthorhombic
symmetry. Earlier reports indicate that the orthorhombic
structure arises from a Bi-O buckling distortion that also affects the
$CuO_{2}$ planes.\cite{Kirk}

In addition to revealing the extended saddle point van Hove singularity,
our data allow us to comment on the possibility of an electronic
state just above the Fermi energy in the $\Gamma-\bar{M}-Z$ direction.
Recently, Ref.\ \onlinecite{Dessau} argued that on optimally doped
$Bi_{2}Sr_{2}Ca_{1}Cu_{2}O_{x}$ samples there are two
closely spaced electronic states, one of even and the other of mixed symmetry.
Along the $\Gamma-\bar{M}-Z$ direction and with even
symmetry orientation, our data are, for the wavevector domain in common,
similar to those reported by Ref.\ \onlinecite{Dessau}. However, as
illustrated in Fig.\ \ref{GammaMZ-vertical}, along the
$\Gamma-\bar{M}-Z$ direction and with odd symmetry orientation, we find
virtually no electronic state.

To insure that this is not
due to a matrix element effect, we took data at several photon energies
in the 17-30 eV range. The difference noted in Fig.\ \ref{GammaMZ-vertical}
is independent of the photon energy.

If we define a polarization factor
$P = (S_{\theta=22^\circ}-S_{\phi=22^\circ})/(S_{\theta=22^\circ}+
S_{\phi=22^\circ})$, where $S_{\theta}$ and $S_{\phi}$ represent the
photoemission spectral area at $\theta$ and $\phi$ after the subtraction
of normal emission spectral area, respectively, the data indicate that for the
overdoped samples, the electronic state along the $\Gamma-\bar{M}-Z$ direction
is $P = 66\%$ of even symmetry, as indicated in
Fig.\ \ref{GammaMZ-vertical}(b).
This result contrasts to that of Ref.\ \onlinecite{Dessau} who, for lighter
doped samples,
report an electronic state in the odd symmetry orientation
that disperses through the Fermi surface along the $\Gamma-\bar{M}-Z$
direction.

The two data sets indicate that the oxygen doping affects the symmetry
of the quasiparticle state along the $\Gamma-\bar{M}-Z$ direction. Our results,
illustrated in Fig.\ \ref{GammaMZ-vertical} and repeatedly reproduced for
overdoped samples, establish that the quasiparticle state exhibits almost
perfect even symmetry in the $\Gamma-\bar{M}-Z$ direction. We emphasize
that our data for
overdoped samples do not allow us to confirm or to exclude the possibility of
a mixed-symmetry state, observable in odd symmetry orientation, that
disperses through the Fermi surface. We simply do not observe strong state in
the odd symmetry orientation.

\section{Conclusion:}
We have mapped the quasiparticle dispersion along the $\Gamma-\bar{M}-Z$
direction up to nearly the Z point for the first time. We find an extended
saddle point type of singularity in the Fermi surface, which can lead to a
stronger than logarithmic enhancement of $T_{c}$. It is noteworthy, as
Abrikosov has pointed out, that this enhancement is virtually independent of
the choice of exchange boson. We have found that the overdoped
$Bi_{2}Sr_{2}Ca_{1}Cu_{2}O_{8+x}$ system exhibits an orthorhombic, rather than
a tetragonal, Fermi surface.

\vspace{.8in}
\noindent ACKNOWLEGEMENT
We benefitted from conversations with Andrey Chubukov and Robert Joynt.
The staff at the Wisconsin Synchrotron Radiation Center, particularly
Tom Baraniak, were most helpful. Financial support was provided by the U.S.
NSF, both directly (DMR-9214701)  and through support of the SRC, by Ecole
Polytechnique F\'{e}d\'{e}rale Lausanne and the Fonds National Suisse de la
Recherche Scientifique, and by the Deutsche Forschungsgemeinschaft.

\begin{figure}
\caption{AC susceptibility data for an oxygen overdoped
$Bi_{2}Sr_{2}Ca_{1}Cu_{2}O_{8+x}$ single crystal annealed in 1 atm $O_{2}$
at 530 $^\circ$C for 20 hours. The onset $T_{c}$ is 83K,
with 10-90\% transition width $\Delta T$ = 1.3K.}
\label{AC}
\end{figure}

\begin{figure}
\caption{Normal state (T=95K) angle-resolved photoemission spectra
for an oxygen overdoped $Bi_{2}Sr_{2}Ca_{1}Cu_{2}O_{8+x}$ single crystal
of $T_{c}$ = 83K along the $\Gamma-\bar{M}-Z$ direction in the Brillouin zone.
The $\Gamma-\bar{M}-Z$ direction lies in the photon polarization plane (even
symmetry). Photon energy of (a) 21 eV , (b) 25 eV employed.
(c) Dispersion curve obtained from (a) and (b). Note the dispersing
electronic state that disperses toward the Fermi energy $E_{f}$ away from
$\Gamma$, the extended flat region between 0.75\AA$^{-1}-$1.25\AA$^{-1}$,
and the dispersion away from $E_{f}$ at wavevectors greater than
1.25\AA$^{-1}$. The dotted line indicates the maximum binding energy for
which a photoemission superconducting condensate peak is
observed.[18]}
\label{GammaMZ}
\end{figure}

\begin{figure}
\caption{(a) Normal state (T=95K) angle-resolved photoemission spectra
for an oxygen overdoped $Bi_{2}Sr_{2}Ca_{1}Cu_{2}O_{8+x}$ single crystal of
$T_{c}$ = 83K along the $\Gamma-\bar{M}-Z$ direction. The $\Gamma-\bar{M}-Z$
direction lies in the photon polarization plane (even symmetry). A photon
energy of 25 eV was employed. (b) The dispersion curve obtained from
Fig. 3(a). Positive values of $k_{y}$ correspond to a wavevector
pointed along the
$\bar{M}-Y$ direction.}
\label{X-M-Y}
\end{figure}

\begin{figure}
\caption{(a).Normal state (95K) angle-resolved photoemission spectra for an
oxygen overdoped $Bi_{2}Sr_{2}Ca_{1}Cu_{2}O_{8+x}$ single crystal of
$T_{c}$ along a direction parallel to the $X-\bar{M}-Y$ direction at
$\theta = 18^\circ$. The $\Gamma-\bar{M}-Z$ direction lies in the photon
polarization plane (even symmetry). The photon energy was 25 eV.
(b). The dispersion curve obtained from Fig.4(a). Positive values of $k_{y}$
correspond to a wavevector parallel to the $\bar{M}-Y$ direction.}
\label{cut18}
\end{figure}

\begin{figure}
\caption{(a). Normal state (95K) angle-resolved photoemission spectra for
an oxygen overdoped $Bi_{2}Sr_{2}Ca_{1}Cu_{2}O_{8+x}$ single crystal
$T_{c}$ =83K along a direction parallel to the $X-\bar{M}-Y$ line at
$\theta = 16^\circ$. The $\Gamma-\bar{M}-Z$ direction lies in the photon
polarization plane (even symmetry). The photon energy was 25 eV.
(b). The dispersion curve obtained from Fig. 5(a). Positive values of $k_{y}$
correspond to a wave vector parallel to the $\bar{M}-Y$ direction.}
\label{cut16}
\end{figure}

\begin{figure}
\caption{(a). Normal state (95K) angle-resolved photoemission spectra for an
oxygen overdoped $Bi_{2}Sr_{2}Ca_{1}Cu_{2}O_{8+x}$ single crystal of
$T_{c}$ = 83K along a direction parallel to the $X-\bar{M}-Y$ direction at
$\theta = 14^\circ$. The $\Gamma-\bar{M}-Z$ direction lies in the photon
polarization plane (even symmetry). The photon energy was 25 eV.
(b). The dispersion curve obtained from Fig. 6(a). Positive values of $k_{y}$
correspond to a wavevector parallel to the $\bar{M}-Y$ direction.}
\label{cut14}
\end{figure}

\begin{figure}
\caption{(a) The experimental positions of the Fermi surface crossings near the
$\Gamma-\bar{M}-Z$ direction. The sample exhibits orthorhombic, not tetragonal,
symmetry, in agreement with structural characterization.
The plane containing the c-axis
and the $\Gamma-\bar{M}-Z$ direction is not a plane of reflection symmetry
of the Fermi surface. (b) The perspective drawing of the binding energy of
the spectral features near the $\Gamma-\bar{M}-Z$ direction. Note that the
band disperses through the Fermi energy as one moves away from the $\Gamma-\bar
{M}-Z$ line in either perpendicular direction, as expected for a extended
saddle
point van Hove singularity.}
\label{FS}
\end{figure}

\begin{figure}
\caption{(a). Normal state (T = 95K) angle-resolved photoemission spectra for
an oxygen overdoped $Bi_{2}Sr_{2}Ca_{1}Cu_{2}O_{8+x}$ single crystal
of $T_{c}$ = 83K along the $\Gamma-\bar{M}-Z$ direction in the Brilouin zone.
The $\Gamma-\bar{M}-Z$ direction lies  perpendicular to the photon
polarization plane (odd symmetry). The photon energy was 21 eV.
Note the loss of spectral intensity compared to Fig. 2(a).
(b). Direct comparison of the spectra taken for two
orientations (odd and even). The dotted lines are the spectra taken at normal
emission. All spectra have been normalized above the Fermi energy and at high
binding energy.}
\label{GammaMZ-vertical}
\end{figure}
\end{document}